\documentclass{vgtc}                          

\ifpdf
  \pdfoutput=1\relax                   
  \usepackage{graphicx}                
  \DeclareGraphicsExtensions{.pdf,.png,.jpg,.jpeg} 
\else
  \usepackage{graphicx}                
  \DeclareGraphicsExtensions{.eps}     
\fi%

\graphicspath{{figures/}{pictures/}{images/}{./}} 

\usepackage{microtype}                 
\PassOptionsToPackage{warn}{textcomp}  
\usepackage{textcomp}                  
\usepackage{mathptmx}                  
\usepackage{times}                     
\usepackage{cite}                      
\usepackage{tabu}                      
\usepackage{booktabs}                  

\onlineid{0}

\vgtccategory{Research}

\vgtcinsertpkg

\title{Data-driven Storytelling in Hybrid Immersive Display Environments}

\author{Xiaoyan Zhou\thanks{e-mail: Xiaoyan.Zhou@colostate.edu}\\ %
        \scriptsize Colorado State University %
\and Yalong Yang\thanks{e-mail: yalong.yang@gatech.edu}\\ %
     \scriptsize Georgia Institute of Technology %
\and Francisco Ortega\thanks{e-mail: fortega@colostate.edu}\\ %
     \scriptsize Colorado State University %
\and Anil Ufuk Batmaz\thanks{e-mail: ufuk.batmaz@concordia.ca}\\ %
     \scriptsize Concordia University %
\and Benjamin Lee\thanks{e-mail: Benjamin.Lee@visus.uni-stuttgart.de}\\ %
     \parbox{1.4in}{\scriptsize \centering University of Stuttgart}}

\teaser{
 \centering
 \includegraphics[width=\linewidth]{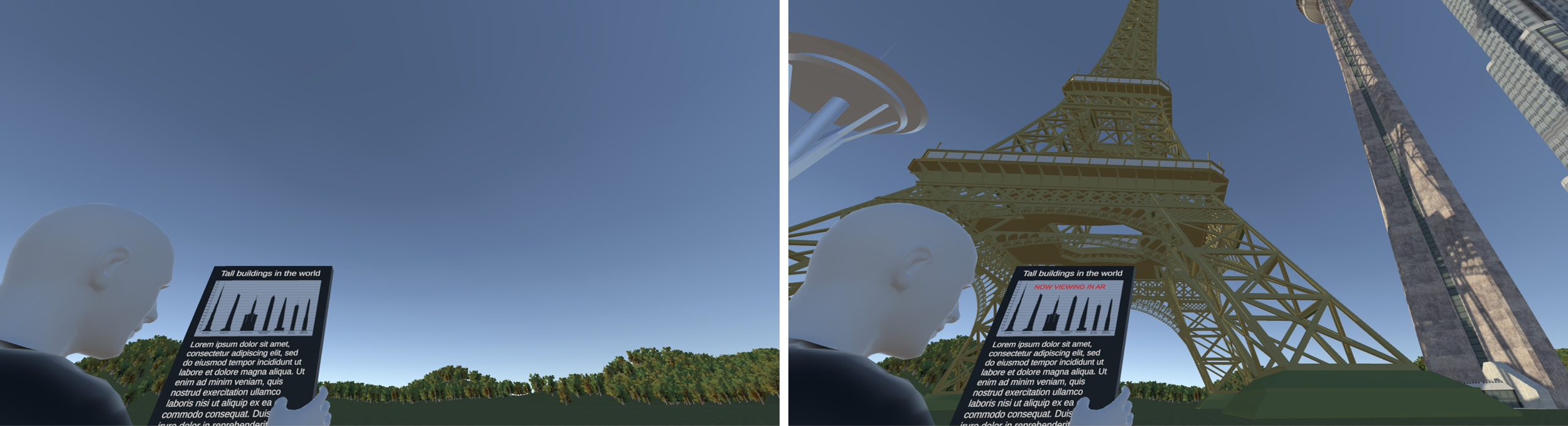}
 \caption{Representative mock-up of a hybrid data-driven story. Left: A person opens and begins reading a regular article on their physical 2D display. Right: They enable a hybrid mode that shows the buildings in the immersive space around them, providing additional context and an accurate portrayal of the buildings' heights while still seeing the original article. Image based on \cite{LEE+20}.}
 \label{fig:teaser}
}

\abstract{Data-driven stories seek to inform and persuade audiences through the use of data visualisations and engaging narratives. These stories have now been highly optimised to be viewed on desktop and mobile computers. In contrast, while immersive virtual and augmented reality (VR/AR) technologies have been shown to be more persuasive, no clear standard has yet emerged for such immersive stories. With this in mind, we propose that a \textit{hybrid} data-driven storytelling approach can leverage the familiarity of 2D display devices with the immersiveness and presence afforded by VR/AR headsets. In this position paper, we characterise hybrid data-driven stories by describing its design opportunities, considerations, and challenges. In particular, we describe how both 2D and 3D display environments can play either complementary or symbiotic roles with each other for the purposes of storytelling. We hope that this work inspires researchers to investigate how hybrid user interfaces may be used for storytelling.}

\begin{document}


\firstsection{Introduction}
\maketitle

Data-driven storytelling, also referred to as narrative visualisation \cite{segelNarrativeVisualizationTelling2010}, is ubiquitous in the media. Journalists, news websites, and even many YouTube channels now combine data visualisation with storytelling techniques to tell compelling, engaging, and persuasive narratives. These come in a range of genres \cite{segelNarrativeVisualizationTelling2010}, ranging from videos \cite{aminiUnderstandingDataVideos2015} to comic strips \cite{BAC+17}. Data-driven stories are commonly web-based and are viewed on 2D displays, with many authoring tools having been developed to support this modality (e.g.~\textit{D3} \cite{bostockDataDrivenDocuments2011}, \textit{SketchStory} \cite{leeSketchStoryTellingMore2013}).

Modern virtual and augmented reality (VR/AR) devices show promise in delivering these data-driven stories. The goals of such immersive technologies do align, after all. Initial attempts towards immersive data-driven storytelling had taken on the form of 360-degree VR videos. However, they suffer from both a lack of true 6 degrees of freedom movement and interactivity---both of which are arguably fundamental to VR/AR. Several works have therefore sought to tell stories using the full capabilities of immersive technologies properly. For instance, Romat et al.~\cite{romatDearPictographInvestigating2020} explored using 3D pictographs in VR to provide more personalised data representations to audiences. Most notably, Lee et al.~\cite{LEE+20} introduced the concept of \textit{data visceralization}, which seeks to facilitate understanding of physical measurements and quantities through visceral experiences in VR. The user can interactively navigate around the virtual environment to ``experience'' the data in different ways, such as to feel runners moving past them at a fast speed. In their work however, they speculated that data visceralization would work best in combination with traditional storytelling techniques: that VR (or even AR) could serve a complementary role to traditional 2D-based stories. Figure~\ref{fig:teaser} displays a representative mock-up of what this may look like.

As per the notion that immersive technologies can serve to enhance our existing tools and workflows in a hybridised setup, we believe that VR/AR can so too enhance traditional data-driven storytelling in many ways.
For example, consider a narrative that aims to convey the effects of climate change (e.g.~\cite{BAL+21, BAL+20, GRE+19}). As the reader progresses through the text-based narrative on their desktop or tablet, they may see their immediate surroundings change to reflect the predicted conditions of the future (e.g.~flooding, droughts), potentially increasing the emotional impact of the story.
Alternatively, consider the scope of interactions that are enabled through the hybrid setup. In a narrative told primarily through AR, the reader may influence the extent to which the effects of climate change are shown by inputting their own CO\textsubscript{2} behaviour on a familiar touch device, thus minimising any overheads in learning new interaction techniques, interfaces, or gestures.
For a third example, consider someone casually watching a news story on their television about rising household energy costs. The story then explains the average energy usage of different household appliances. The viewer's AR headset begins to show embedded views of energy usage data \cite{willettEmbeddedDataRepresentations2017} directly on top of their own personal appliances, thus providing a personalised experience of the narrative.

To the best of our knowledge, no work has considered how data-driven stories can be told and experienced in a hybrid, cross-reality configuration. We speculate that this approach can take full advantage of the existing work on data-driven storytelling by both practitioners and researchers, while also leveraging the strengths of immersive display devices. In this position paper, we identify and discuss some of the key design opportunities enabled through this hybrid setup, followed by the (research) challenges of creating these data stories.

\section{Design Opportunities and Considerations for Hybrid Data-driven Stories} \label{sec:opportunities}
While 2D screens can provide a consistent and reliable narrative (typically through text and 2D graphics), 3D environments offer an entirely novel and vast canvas in which narratives can be told. When considering how the two might interplay with one another in a hybridised setup, it is apparent that the set of design considerations grows rapidly. In this section, we aim to summarise the key considerations that we believe warrant future exploration and discussion.

\subsection{Data and their visual representations}
In data-driven storytelling, it is common to use tools such as D3 \cite{bostockDataDrivenDocuments2011} to create interactive web-based 2D visualisations interwoven with text. Of course, we can just as easily display 3D visualisations in the accompanying immersive environment (e.g.~using \textit{IATK} \cite{cordeilIATKImmersiveAnalytics2019}), whether they be standalone or be juxtaposed against another 2D visualisation in a coordinated view \cite{ROB+22}. However, it is no secret that many 3D visualisations suffer from perceptual issues \cite{graciaNewInsightsSuitability2016}. When there is no suitable 3D representation for the given data, a storyteller may resort to only using 2D visualisations, even as floating windows in the immersive environment.

Of course, we are not limited to just 3D data visualisations---the immersive scene can theoretically render anything the storyteller desires. As mentioned, one such approach is to represent data as virtual objects in order to allow audiences to ``experience'' their data in VR \cite{LEE+20}. A more standard approach is to use this as a form of concrete scales \cite{chevalierUsingConcreteScales2013}, such as to represent water usage per flush in the form of water bottles placed directly next to a toilet \cite{assorExploringAugmentedReality2023}. These examples demonstrate the explicit intention to represent data as other (virtual) objects.

Data need not even be encoded however. Immersive contextual aids can very well be added to existing 2D narratives that do not explicitly serve to showcase data. For instance, a text narrative about COVID-19 deaths may situate the reader into a virtual recreation of a hospital, much in the same way a video might show B-roll stock footage for additional context. While extra embellishments which surround the data are oftentimes admonished as a form of chart junk in traditional information visualisation \cite{tufteVisualDisplayQuantitative2013}, they can possibly serve to actually benefit the persuasiveness and enjoyment of data-driven stories.

\subsection{Managing layout of content and narrative}
Segel and Heer~\cite{segelNarrativeVisualizationTelling2010} initially identified seven genres of data-driven storytelling, which represent the most commonly used orderings and layouts to tell stories with. Such layouts matter when telling a story, as they influence whether the story is highly linear or if the reader is free to choose their own order. A now very popular form of data-driven storytelling is that of \textit{scrollytelling}~\cite{seyserScrollytellingAnalysisVisual2018}: a long-form type of narrative that relies heavily on text interwoven with multimedia that is read from top to bottom and facilitated via scrolling (hence the name). While this provides little freedom to the reader, it allows the author to carefully construct and drive the narrative.

In contrast, the ways in which the content and narrative can be organised in immersive environments are much greater---even when considering the immersive environment by itself. Judging by the myriad works on layout and view management in immersive environments (e.g.,~\cite{luoWhereShouldWe2022,leeSharedSurfacesSpaces2021}), a simple approach would be to position 2D panels in space. This alone, we hypothesise, would not lead to an engaging experience, at least without some additional form of visual stimulus or embellishment alongside said panels. Other approaches may prove beneficial, such as leveraging the spatiality of VR/AR to have the reader move through a 3D space as a form of navigation throughout the story (akin to \textit{DataHop} \cite{hayatpurDataHopSpatialData2020}). Depending on the layout and design of these highly spatial narratives however, some form of attention guidance may be required to properly guide the reader through each step of the story (e.g.~\cite{danieauAttentionGuidanceImmersive2017}).

When now used in a hybrid setup, the consideration of content layout is important. Content can be shown primarily on the 2D display or on the 3D display, essentially dictating which one serves as the main driver for the narrative, with the other playing a complementary role (akin to \cite{zhuBISHAREExploringBidirectional2020}). Alternatively, both displays may be equally important, with the narrative moving between the two in sequential order.

For a \textbf{2D display-centric} narrative, immersive 3D content would likely need to be placed nearby or around said display. This approach would serve to easily complement existing 2D data-driven stories with the advantages of immersive devices. This may be to provide alternate 3D views of the data, or to provide additional context by displaying relevant objects or scenes around the reader. Important to consider, however, is the fine line between VR/AR being complementary and it being distracting. A 3D embellishment that is too visually salient may divert attention from the main narrative, disrupting its flow. On the other hand, important 3D visuals that are not clearly visible (e.g.~out of view) may prove essentially worthless, or even cause confusion if imperative to the narrative. Thus, techniques would likely need to be employed to ensure the reader's attention is properly redirected were necessary, without it being too overt or annoying.

For a \textbf{3D display-centric} narrative, the 2D display may serve different purposes. In some ways, a 2D display is redundant in an immersive environment as virtual 2D panels can be created at will. However, we argue that conventional computing devices such as desktops and tablets serve to provide a familiar and tangible interface that is not otherwise possible with a pure VR/AR setup. For instance, they might act as a familiar landmark to display text on that one can readily access, rather than the reader needing to locate a floating text panel whose adaptive positioning rules may differ from application to application. They might also serve as a tangible input device, whether to input data values (e.g.~in service of a personalised narrative) or to control and interact with the 3D environment (e.g.~\cite{serenoSupportingVolumetricData2019}). They can even serve to simply facilitate the browsing and loading of hybrid data-driven stories to begin with, using a familiar touchscreen rather than a potentially complex and fatiguing spatial user interface.

\subsection{Transitions between environments}
In the case of \textbf{equivalent} narratives between 2D and 3D, the act of transitioning between the two and the method by which it is done is important. As the reader switches between the two, they incur a form of context-switching cost which may disrupt the narrative. For instance, if the layout of the 3D environment is vastly different from the 2D environment, readers may feel disoriented and require time to adjust. This challenge is not unique to hybrid data-driven storytelling, as it is also present in cross-virtuality analytics applications \cite{HUB+22,frohlerSurveyCrossVirtualityAnalytics2022}. That said, in the context of data-driven storytelling, possible solutions arise which may reduce this cost. Most notably, an embodied narrator can guide the reader as the story moves between environments. Through this, the switching cost may be minimised as the context has shifted on the singular narrator itself, which may then assist in reorienting the reader through careful attention guidance. Of course, it would be prudent also to explore other possible solutions to this issue.

It is also important to consider when the transition between the two environments actually occurs. In the aforementioned case of an embodied narrator (or even without), the transition may occur when dictated by the story (i.e.~chosen by the author). In this situation, the reader does not need to actively consider what purpose either environment serves---they simply follow the narrative. In contrast, the choice to transition may be instead decided by the reader themselves. In this situation, they would need to be aware of the purpose of both environments, and make a conscious decision whether they want to switch. For example, a reader might want to switch from the 3D to 2D environment when they have to read a lot of text, or perhaps when they are tired and would rather be seated and lean back. Alternatively, they might want to switch from the 2D to 3D environment when they would like to see a more immersive representation of the data \cite{LEE+20}. That said, providing this freedom necessitates a \textbf{parallel} narrative in both 2D and 3D environments. The narrative needs to be fully complete, cohesive, and coherent in both modes, and the transition should be capable of preserving context and narrative timing across them. This includes both the data visualisations used and the messaging techniques employed.

\subsection{Interacting with the narrative}
Some existing genres of stories, such as data videos, allow little to no interaction---only to play, pause, and so on. Others, such as drill-down stories \cite{segelNarrativeVisualizationTelling2010}, encourage readers to choose what parts of the story interest them the most and to dig deeper, and generally use a selection mechanism to do so (such as mouse input). For data visualisations, some interactions may also be allowed, such as brushing and linking across multiple coordinated views \cite{bujaInteractiveDataVisualization1991} and/or changing the view specification via filtering, sorting, changing encodings, etc.~\cite{heerInteractiveDynamicsVisual2012}.

Of course, any interaction technique used on 2D displays would naturally be usable in a hybrid setup, provided that the same display and interaction modalities are present. Not only can ``traditional'' interaction modalities like pointing devices and touch input be used to interact with visualisations on the 2D display, but they may also be used to interact with the visualisations on the 3D display. While these techniques would likely be similar to those already created for cross-virtuality analytics (e.g.~mouse input \cite{wangUnderstandingAugmentedReality2020}, touch input \cite{hubenschmidSTREAMExploringCombination2021}), interactions would not just be used to manipulate visualisations in the 3D world. The narrative itself can also be influenced based on this user input. This influence may either be implicit through basic navigation actions (e.g.~scrolling), or be explicit with the reader directly choosing what the story shows next (e.g.~in a drill-down story). Other objects, such as the aforementioned 3D embellishments, could also be indirectly interacted with as well. As a basic example, consider toggleable modes which adjust their visibility and/or appearance.

In contrast, the possibilities for interactions facilitated by the immersive display are vast. Once again, these post-WIMP techniques can be used to directly manipulate and control visualisations (or other virtual objects) in the 3D display, as per immersive analytics \cite{buschelInteractionImmersiveAnalytics2018}. This obviously depends on the interaction modalities that are actively being used, whether it be embodied interaction via hand gestures, or even voice commands used to engage in a conversational dialogue with the narrator or narrative. A more subtle form of interaction is that of navigation and movement. For interacting with data visualisations, this can be seen as a form of proxemics \cite{ballendatProxemicInteractionDesigning2010}. When we consider how navigation can influence the narrative however, the reader's position within the 3D environment may serve to direct and move through the narrative. For example, imagine a series of virtual 3D exhibits which are arranged in a linear order. As the reader moves close to each exhibit, their handheld 2D device shows detailed information about said exhibit. In this case, navigation serves to progress between stages of the narrative. This does depend on the spatial environment of the 3D environment of course. A more evenly distributed set of virtual exhibits would provide a more open-ended, reader-controllable form of narrative. Alternatively, a story in the form of a roller coaster would result in a linear narrative which the reader can't directly control---they are instead (literally) along for the ride \cite{casamayouRideYourData2022}. Other forms of 3D interaction may also in turn influence the 2D content. The aforementioned example adjusts the text displayed on a handheld 2D device based on 3D movement. As another example, consider narratives in which the reader is tasked in using a hand gesture to initiate a transition between devices, perhaps to ``bring'' some important data out into the 3D world (similar to \cite{leeDesignSpaceData2022}).

Problems may arise when the interaction coupling between 2D and 3D displays are high, but their effects are spatially far apart. A 2D slider on a desktop that adjusts the amount of money visualised as \$100 notes \cite{LEE+20} may be perceived as non-functional if the stacks of money are behind the reader. In these types of situations, some form of attention guidance could be used to properly inform and direct the reader to look at the relevant areas. Alternatively, the narrative layout could be adjusted to ensure the money stacks are always within the reader's field of view.

\section{Challenges of Hybrid Data-driven Stories}
The context and usage of hybrid data-driven storytelling differ greatly from that of not only traditional approaches, but also immersive (non-hybrid) data-driven storytelling itself. We now discuss relevant challenges that exist outside of the realm of narrative design.

\subsection{Authoring hybrid data-driven stories}
While the design of hybrid data-driven stories involves many challenges (and opportunities), the same goes for their authoring and creation. Although tools exist to help creators sketch and create spatially aware hybrid spaces (e.g.~\cite{jetterVREverythingPossible2020}), no tools are presently available to create hybrid stories. Even so, the tools for creating purely immersive stories that do not incorporate other devices are limited (e.g.,~\cite{renXRCreatorInteractiveConstruction2018}). Toolkits such as IATK \cite{cordeilIATKImmersiveAnalytics2019} need to be repurposed and adapted for this new context, presenting significant overheads in implementation. The means in which we can effectively create hybrid data-driven stories remains an open challenge for researchers and practitioners alike. In contrast, it is now easier than ever to author web-based stories thanks to visualisation toolkits like D3 \cite{bostockDataDrivenDocuments2011} and interaction toolkits like React. As we view hybrid stories as potentially being extensions to existing 2D stories, it may be prudent to leverage this existing capability and establish interoperability between 2D web pages and 3D environments, such as by using WebXR\footnote{\url{https://immersiveweb.dev/}}. Authoring is potentially made harder by the potential need to design and develop 3D models and environments. Whereas white space may be acceptable on a 2D display, large spans of empty 3D space with visualisations and information randomly scattered throughout may seem odd or unsightly.

\subsection{Bystander and context awareness}
At present, many data-driven stories can comfortably be read on any device capable of web browsing. The circumstances in which these stories are read typically do not need to be considered by storytellers, whether in private or public. For immersive data-driven storytelling, it is easy to assume the reader is in a large private area. This is, of course, not always the case. While it is considered normal to read a news article or data-driven story on your phone in public, reading the same story while wearing a VR/AR device may be awkward \cite{bajorunaiteVirtualRealityTransit2021} or even dangerous, depending on the level of presence targeted by the author. Previous work has suggested that designers should aim to achieve ``optimal presence'' rather than ``maximal presence'' \cite{RIV+04, GEO+20}. This results in a balance between immersion in the story and awareness of any bystanders and obstacles in the reader's physical environment. Research has suggested techniques for improving bystander awareness, such as using ``peeking windows'' to see into the real world \cite{WIL+19}, or to show bystanders as avatars or widgets \cite{kudoBalancingVRImmersion2021}.

Even so, we speculate that it is more important to adjust the design of the narrative experience itself, as the use of awareness aids may either be distracting or insufficient in certain scenarios. For instance, awareness of bystanders is useless if the narrative requires the reader to physically move in a crowded area. Thus hybrid data-driven stories may need to accommodate the context that the reader is in. Consider someone reading said story on their phone with an AR headset. The narrative may shift to primarily use the 2D display, with 3D embellishments being added in the space directly in front of them. This allows them to take advantage of hybrid data-driven stories even in physically constrained environments. The notion of context-aware technologies is in itself not new \cite{baldaufSurveyContextawareSystems2007}, but may prove invaluable for hybrid data-driven storytelling (or even regular immersive storytelling) to flourish.

This context awareness extends beyond location and environmental factors, but also technological factors. A wide number of hybrid configurations are theoretically possible, for both 2D displays (desktop, phone, tablet, etc.) and 3D displays (optical vs video see-through AR, VR). Therefore, a hybrid data story should ideally adapt to different configurations, akin to web browsers adapting to variations in screen sizes, aspect ratios, and interaction modalities.

\subsection{Evaluating hybrid data-driven stories}
When evaluating data-driven stories, it is common to use measures such as subjective engagement, enjoyment, satisfaction, or even willingness to change one's own behaviour \cite{fauvilleVirtualRealityPromising2020}. In a hybrid context, this is likely to be no different. A key difference however is that hybrid data-driven stories can be decomposed into individual components, each in service to a different goal. Lee et al.~\cite{LEE+20} noted that while ``visceralized'' 3D representations of data may not serve well to convey exact numerical measures and quantities, it ultimately achieves the goal of being more entertaining and emotionally impactful to readers. They also suggested that combining data visceralizations with traditional text-based narratives may serve to cover any factual gaps. In this sense, it is apparent that there may sometimes be two different yet complementary goals of the hybrid setup. An evaluation may choose to isolate each component to determine how well it achieves its goals, and seek to identify any problems that arise. On the other hand, it is likely that, depending on the level of coupling between the two environments, any form of separation becomes impossible, and the entire narrative needs to be evaluated as a whole.

\section{Conclusion}
This position paper proposes the concept of hybrid data-driven storytelling, which seeks to engage, entertain, educate, and persuade readers through the combination of conventional 2D displays and immersive 3D devices. A core advantage of the approach is its ability to extend existing stories with an immersive component, whether it be to provide 3D views of data, provide additional context, or to increase enjoyment. We hope that our work not only demonstrates the interesting and exciting possibilities of hybrid data-driven storytelling, but also sets out clear goals for future research.


\acknowledgments{
This research was funded by the German Research Foundation (DFG) project 495135767 and the Austria Science Fund (FWF) project I 5912-N (joint Weave project).}

\bibliographystyle{abbrv-doi}

\bibliography{template}
\end{document}